# Perfect drain for the Maxwell Fish Eye lens

**Juan C. González, Pablo Benítez, and Juan C. Miñano**

Universidad Politécnica de Madrid, Cedint
Campus de Montegancedo
28223 Madrid, Spain

E-mail: jcgonzalez@cedint.upm.es

**Abstract.** Perfect imaging for electromagnetic waves using the Maxwell Fish Eye (MFE) requires a new concept: the perfect drain. From the mathematical point of view, a perfect point drain is just like an ideal point source, except that it drains power from the electromagnetic field instead of generating it. We show here that the perfect drain for the MFE can be seen as a dissipative region the diameter of which tends to zero. The complex permittivity ε of this region cannot take arbitrary values, however, since it depends on the size of the drain as well as on the frequency. This interpretation of the perfect drain connects well with central concepts of electromagnetic theory. This opens up both the modeling in computer simulations and the experimental verification of the perfect drain.

## 1. Introduction

The possibility of focusing light below the diffraction limit (super-resolution) has been demonstrated in the last decade using left-handle materials [1][2][3][4] (that is, materials with negative dielectric and magnetic constants [5]). Recently, a new possibility has been introduced using the Maxwell Fish Eye (MFE) lens, for a material with a positive, isotropic and inhomogeneous refractive index. It is well known that, in the Geometrical Optics (GO) framework, the MFE perfectly focuses rays emitted by an arbitrary point of space into another (its image point). Leonhardt [6] has demonstrated in that the MFE lens in two dimensions not only perfectly focuses radiation in the GO approximation, but also does so for actual fields of any frequency, a result that has been confirmed via a different approach [7]. This two dimensional analysis describes TE-polarized light in a cylindrical medium (where the electric-field vector E points orthogonal to the plane), and the electric field magnitude fulfills the Helmholtz equation. Leonhardt and Philbin have also demonstrated the analogous ideality of a novel impedance-matched spherical MFE for perfect focusing of electromagnetic waves in three dimensions [8].

In the two-dimensional case, the perfect focusing of the MFE in [6] assures that the medium will perfectly transport an outward (monopole) Helmholtz wave field, one generated by a point source, towards an “infinitely-well localized drain” [6] (one that we will call “perfect point drain”) located at the desired image point. Note that the perfect point drain must be such that it totally absorbs all incident radiation, with no reflection or scattering by it. Note also that the field around the drain asymptotically coincides with an inward (monopole) wave. We will refer here to such a wave as “Leonhardt’s forward wave”.

Even though the physical significance of a point source as a limiting case seems to be well accepted, that of a passive perfect point drain has been considered very controversial [9][10][11][12]. In reference [6], the drain was not physically described, but only considered as a mathematical entity, leaving no clues as to how to simulate that drain in software. Particularly, an analysis of such a drain, one located at a position different from the image point, would help to prove the super-resolution, which could not be done with the information in reference [6].

Recently, however, a candidate for perfect drain has been proposed for a microwave-frequency MFE [13], wherein a two-dimensional MFE medium has been assembled as a planar waveguide with concentric layers of copper circuit board forming the desired refractive index profile of the MFE. Also, both source and drain have been built as identical coaxial probes, one to introduce power into the planar waveguide and the other to extract it. The coaxial probe was intended to act as the perfect sink and is completely passive and loaded with the characteristic impedance of the coaxial cable [13]. There is no theoretical proof that such a coaxial probe will actually behave as a perfect drain, but a practical proof is claimed by comparing (Figure 4 in [13]) the measured electric field distribution and the analytical expression of Leonhardt's forward wave ([6], reviewed here in Section 1.1). Since the measured and analytical values differ significantly (deviations attributed primarily to imperfections in the probes), we think a more detailed analysis is called for in order to clarify whether or not that receiving coaxial probe is acting as the perfect drain as defined here. Nevertheless, this clarification does not affect the main conclusion in [13]: such a probe certainly leads to super-resolution in the conditions of that measurement.

In this paper we present a different realization of a passive perfect drain, obtained theoretically from the analytical equations. It consists in the introduction of a certain non-magnetic material inside a circle of radius $R$ enclosing the image point (although not centered upon it, as discussed later). As will be proved in Section 2, the complex permittivity of that non-magnetic material will be such that the field outside that circle coincides exactly with Leonhardt's forward wave and the incoming power will be fully absorbed inside it. In general, this perfect drain can have a finite radius $R$; the perfect point drain is just the limit case $R\rightarrow 0$.

### *1.1 Leonhardt's forward wave*

The strength of the TE monochromatic field $E(x,y)e^{-i\omega t}\mathbf{z}$ in a region without sources or drains fulfils the 2D-Helmholtz equation:

$$\Delta E + n^2 k_0^2 E = 0 \tag{1}$$

where $k_0 = \omega\sqrt{\mu_0\varepsilon_0}$ and $n$ is the MFE refractive index distribution given by:

$$n = \frac{2}{1+\rho^2} \tag{2}$$

where $\rho = \sqrt{x^2+y^2}$ . Leonhardt's forward wave is a particular family of solutions of Eq. (1), given by Eq. (12) in [6]:

$$E = \frac{P_\nu(\zeta) - e^{i\nu\pi} P_\nu(-\zeta)}{4\sin(\nu\pi)} \tag{3}$$

where $P_\nu$ is the Legendre function of the first kind,

$$\nu = \frac{-1+\sqrt{1+4k_0{}^2}}{2} \tag{4}$$

and

$$\zeta = \frac{|z'|^2 - 1}{|z'|^2 + 1} \qquad z' = \frac{z - x_0}{z x_0 + 1} \tag{5}$$

Here $z=x+iy$ is the complex notation of the point (x,y), and $x_0$ is an arbitrary real number. Without lack of generality we have located the point source described in [5] at the object point ($x_0$, 0). Note that $-1\leq\zeta\leq 1$ and by the divergence of $P_\nu(\zeta)$ when $\zeta\rightarrow -1$, $E$ is infinite at $|\zeta|=1$. A wave according to Eq. (3) is generated by the point source located at ($x_0$,0), and it propagates towards the perfect point drain at the image point ($-1/x_0$, 0). The time evolution of the field, $\mathrm{Re}(E(x,y)e^{-i\omega t})$ is shown in the associated media files Media1a and Media1b for $x_0=-2$, $k=15$. This evolution clearly includes the one-directional propagation of the wave from source to drain, with no reflection or scattering by it.

Note that the magnetic field **H**(x,y) of Leonhardt's forward wave can be easily computed from the field **E**, Eq.(3), and Eq. (5), as:

$$\mathbf{H}=\frac{1}{i\omega\mu}\nabla x\mathbf{E}=\frac{1}{i\omega\mu}\frac{dE}{d\zeta}\left(\frac{\partial\zeta}{\partial y}\mathbf{x}-\frac{\partial\zeta}{\partial x}\mathbf{y}\right) \tag{6}$$

### *1.2 Current of the source*

Eq. (1) is only valid in a region without sources or drains, in our case the full plane except for the points ($x_0$, 0) and ($-1/x_0$, 0). The equation valid for the full plane must include the Dirac delta at those points. The amplitude of Eq. (3) was specifically selected in [6] to make it behave as a Green function, i.e., so the weights of those Dirac deltas have unit moduli. It can be written as [6][7][8]:

$$\Delta E+n^2k_0^2E=-\delta(x-x_0,y)-\mathrm{e}^{i\nu\pi}\,\delta(x-1/x_0,y) \tag{7}$$

The right hand side of Eq. (7) can be identified (from Maxwell equations) as $-i\omega\mu_0 J(x,y)\mathbf{z}$, where $J(x,y)$ is the current density. Consequently the electric currents through source and drain for the specific field amplitude of Eq. (3) are:

$$I_{source}=\frac{1}{i\omega\mu_0}\qquad I_{drain}=\frac{e^{i\pi\nu}}{i\omega\mu_0}=e^{i\pi\nu}I_{source} \tag{8}$$

### *1.3 Alternative expression for the forward wave*

There is an alternative way to express Eq. (3) (Eq. 12 of [6]), which uses the Legendre function of the second kind $Q_\nu$. We consider here the branch of $Q_\nu$ that is real-valued when $\mathrm{Im}(\zeta)=0$ and $|\zeta|<1$ (another complex-valued branch $Q_\nu$ is claimed to be considered in [6]). Taking into account, from [14] (Eq. (15) in p. 144), that:

$$P_\nu(-\zeta)=\cos(\pi\nu)P_\nu(\zeta)-\sin(\pi\nu)\frac{2}{\pi}Q_\nu(\zeta) \tag{9}$$

We already know that the Eq. (3)can be alternatively rewritten as:

$$E=-ie^{i\pi\nu}\left(P_\nu(\zeta)+i\frac{2}{\pi}Q_\nu(\zeta)\right) \tag{10}$$

Let us designate the factor inside the parenthesis of (10) as:

$$F_\nu(\zeta)=P_\nu(\zeta)+i\frac{2}{\pi}Q_\nu(\zeta) \tag{11}$$

Interesting asymptotic expressions for function $F_\nu$ are set forth in Appendix 1, which gives comprehensive discussion of Leonhardt's forward wave. Additionally, this alternative expression will greatly simplify some calculations in the perfect sink design of the next section.

## 2 Perfect sink

The theoretically ideal point drain of Leonhardt is located at the point ($-1/x_0$, 0). We will design first a finite-area perfect drain, which will comprise a convex region surrounding that point ($-1/x_0$, 0), filled with an inhomogeneous, isotropic, non-magnetic material with complex dielectric permittivity (thus, absorptive), such that the field outside that region coincides exactly with Leonhardt's forward wave.

The equation satisfied by the field in that region will then be the homogeneous Helmholtz equation. Since the incident wave fields **E** and **H** are known, the necessary continuity of **E** and **H** on the drain boundary will be forced by particularizing on the boundary the values of **E** and **H** in Eq. (3) and Eq.(6), respectively.

### *2.1 Selection of the boundary*

From Eq. (5), it is easy to confirm that the line $\zeta=\zeta_d$ =const is a circle with its centre at the point ($x_c$, 0) and its radius $R$ given by:

$$x_c = \frac{2x_0}{1-x_0^2-\left(1+x_0^2\right)\zeta_d} \qquad R = \frac{-\sqrt{1-\zeta_d^2}\left(1+x_0^2\right)}{1-x_0^2-\left(1+x_0^2\right)\zeta_d} \tag{12}$$

We select the drain region as that containing the points fulfilling $\zeta \geq \zeta_d$, with $\zeta_d$ fulfilling:

$$\zeta_d > \frac{1-|x_0|^2}{1+|x_0|^2} \tag{13}$$

Condition Eq. (13) is required in order to guarantee that the drain area is finite and encloses the image point ($-1/x_0$, 0), as can be deduced from Eq. (12). The $\zeta$=constant circumferences are shown in red in Fig.1. The $\zeta$=constant circumferences are also $E$=constant lines (see Eq.(3)), and coincide with the GO wavefronts. This property is fulfilled by a more general class of inhomogeneous media, a class analyzed in [7]. The blue lines in Fig.1, which we can call $\delta$=constant, are also circumferences, and coincide with the GO rays. and are the Poynting vector lines of Leonhardt's forward wave. Both families of lines coincide with coordinate grid lines of the bipolar orthogonal coordinate system (see, for instance, [7]).

From Eq. (12) we see that selecting $\zeta_d$ close enough to 1 will make the radius $R$ as small as desired. At the limit $\zeta_d \rightarrow 1$, we obtain the point drain ($R \rightarrow 0$, $x_c \rightarrow -1/x_0$).

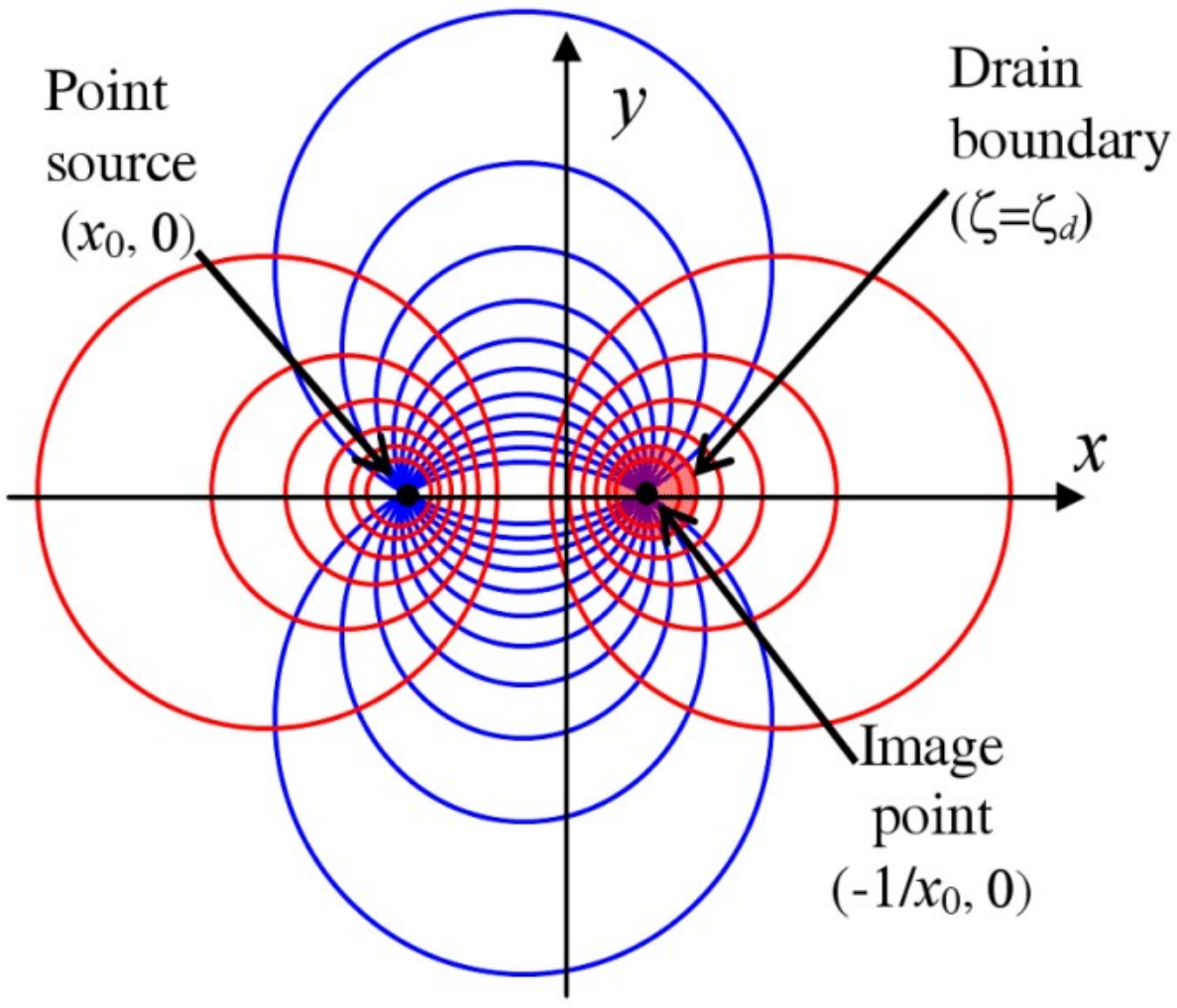

Fig. 1. The source is a line of current perpendicular to the x-y plane and placed at the point ($x_0$, 0), which defines the two-dimensional point source. The boundary of the finite-area passive drain is a cylinder with base $\zeta > \zeta_d$. The red lines ($\zeta$=constant) and the blue lines ($\delta$=constant) define a cylindrical bipolar coordinate system.

### *2.2 Inhomogeneous complex refractive index of the drain*

Inside the drain ($\zeta \geq \zeta_d$), we select $\mu = \mu_0$ and the refractive index has the following form:

$$n_d = \frac{2\sqrt{\varepsilon_d}}{1+\rho^2} \tag{14}$$

where $\varepsilon_d$ is a complex constant and with Im($\varepsilon_d$)≠0 to be calculated later (in Section 3). Then, the homogeneous Helmholtz equation in the drain is:

$$\Delta E + n_d^2 k_0^2 E = 0 \tag{15}$$

Designate $k_d = \omega\sqrt{\mu_0 \varepsilon_0 \varepsilon_d}$ (which is complex). Using the expression for the refractive index $n$ of the MFE (Eq. (2)), we find that the selection made with Eq. (14) fulfills that $n_d k_0 = n k_d$, so Eq. (15) can also be written as:

$$\Delta E + n^2 k_d^2 E = 0 \tag{16}$$

This equation is identical to the Helmholtz equation of the MFE, Eq.(1), after substituting the real wave number $k_0$ by $k_d$ (still to be calculated).

### *2.3 Ordinary differential equation of the drain*

One of the boundary conditions on the line $\zeta = \zeta_d$ is the continuity of field **E**=$E$**z**. As said before, $E$ is constant on that boundary surface, so it would be particularly interesting to express Eq. (16) in the bipolar coordinate system $\zeta$-$\delta$. That was already done in Section 3.1 of reference [7]. As shown there, the expression of Eq. (16) for solutions depending only on $\zeta$ is the same as that Eq. (9) in [6]. Using the change of variables $\zeta=(r^2-1)/(r^2+1)$, this equation is:

$$\frac{1}{r}\frac{d}{dr}\left(r\frac{dE}{dr}\right)+n^2k_d^2E=0 \tag{17}$$

Its general solution [14] can be written as:

$$E=A\,P_{\nu'}(\zeta)+B\,Q_{\nu'}(\zeta) \tag{18}$$

where $\nu'$ is given by

$$\nu'=\frac{-1+\sqrt{1+4k_d^2}}{2} \tag{19}$$

The three constants $A$, $B$ and $\nu'$ are fixed by three conditions. Two of them are given by the continuity of the **E** and **H** fields at the boundary, and the third is that the field $E$ must be bounded (*i.e.*, it cannot diverge), since the Helmholtz Eq. (15) is homogeneous.

Consider first the third condition. From the properties of the Legendre functions [14], we know that the function $Q_{\nu'}(\zeta)$ diverges when $\zeta\to 1$, if $\mathrm{Im}(\nu')\neq 0$ and $P_{\nu'}(\zeta)$ does not ($P_{\nu'}(1)=1$). Therefore, this boundary condition imposes $B=0$, which means that the field inside the drain region ($\zeta \geq \zeta_d$) has the form:

$$E=A\,P_{\nu'}(\zeta) \tag{20}$$

$A$ and $\nu'$ are calculated forcing the other two conditions i.e., the continuity of **E** and **H** at the boundary. **E** and **H** outside the drain are taken from the solution in the absence of reversed wave (so the drain is reflection-less) *i.e.*, by using Eq. (10) and Eq. (6)).

$$\begin{aligned} A\,P_{\nu'}(\zeta_d) &= -ie^{i\pi\nu}\left(P_\nu(\zeta_d)+i\frac{2}{\pi}Q_\nu(\zeta_d)\right) \\ A\left.\frac{dP_{\nu'}}{d\zeta}\right|_{\zeta_d} &= -ie^{i\pi\nu}\left(\left.\frac{dP_\nu}{d\zeta}\right|_{\zeta_d}+i\frac{2}{\pi}\left.\frac{dQ_\nu}{d\zeta}\right|_{\zeta_d}\right) \end{aligned} \tag{21}$$

Dividing both equations, we obtain:

$$\frac{\left.\dfrac{dP_{\nu'}}{d\zeta}\right|_{\zeta_d}}{P_{\nu'}(\zeta_d)}=\frac{\left.\dfrac{dP_\nu}{d\zeta}\right|_{\zeta_d}+i\dfrac{2}{\pi}\left.\dfrac{dQ_\nu}{d\zeta}\right|_{\zeta_d}}{P_\nu(\zeta_d)+i\dfrac{2}{\pi}Q_\nu(\zeta_d)} \tag{22}$$

which is an expression where only $\nu'$ is unknown. Once this is solved we can calculate $\varepsilon_d$ with Eq. (19) :

$$\varepsilon_d=\left(\frac{k_d}{\omega\sqrt{\mu_0\varepsilon_0}}\right)^2=\frac{\nu'(\nu'+1)}{\omega^2\mu_0\varepsilon_0} \tag{23}$$

### 3. Examples

Fig. 2 shows $\varepsilon_d$ as function of frequency for two different drain radii and Fig. 3 shows it as a function of the drain radius for different frequencies. In all the examples we shall assume that $\rho$ in Eq (2) and (14) is given in cm and frequency in GHz. Media2a.mov and Media2b.mov show the time evolution of the field, $\mathrm{Re}(E(x,y)e^{-i\omega t})$ for a drain radius $R$=0.2cm, $x_0$=-2, $k$=15, This should be compared with the case of the perfect point drain shown in the Introduction (Media2a and Media2b). Media3 shows the time evolution of the field inside the drain for radius 0.2 cm.

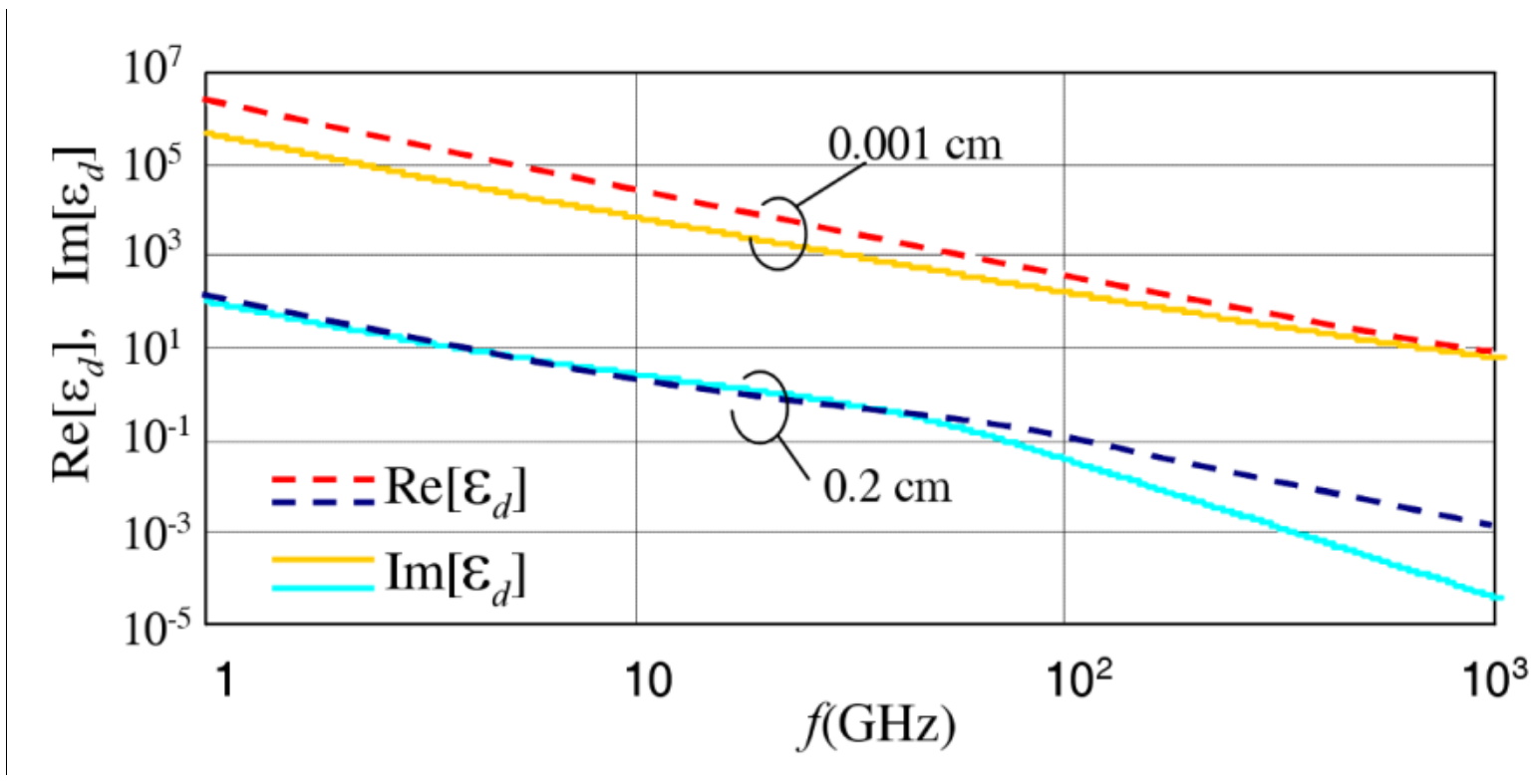

Fig. 2. Dependence of $\varepsilon_d$ (Eq. (14) with frequency. Dark and bright blue curves correspond to the real and imaginary part for a drain radius 0.2 cm centered at $x_c$=0.5159 cm, $y_c$=0. Red and orange curves correspond to the real and imaginary parts for a drain radius 0.001 cm centered at $x_c$=0.5 cm, $y_c$=0. The point source is located at $x_0$=-2 cm, $y_0$=0. Frequency in GHz.

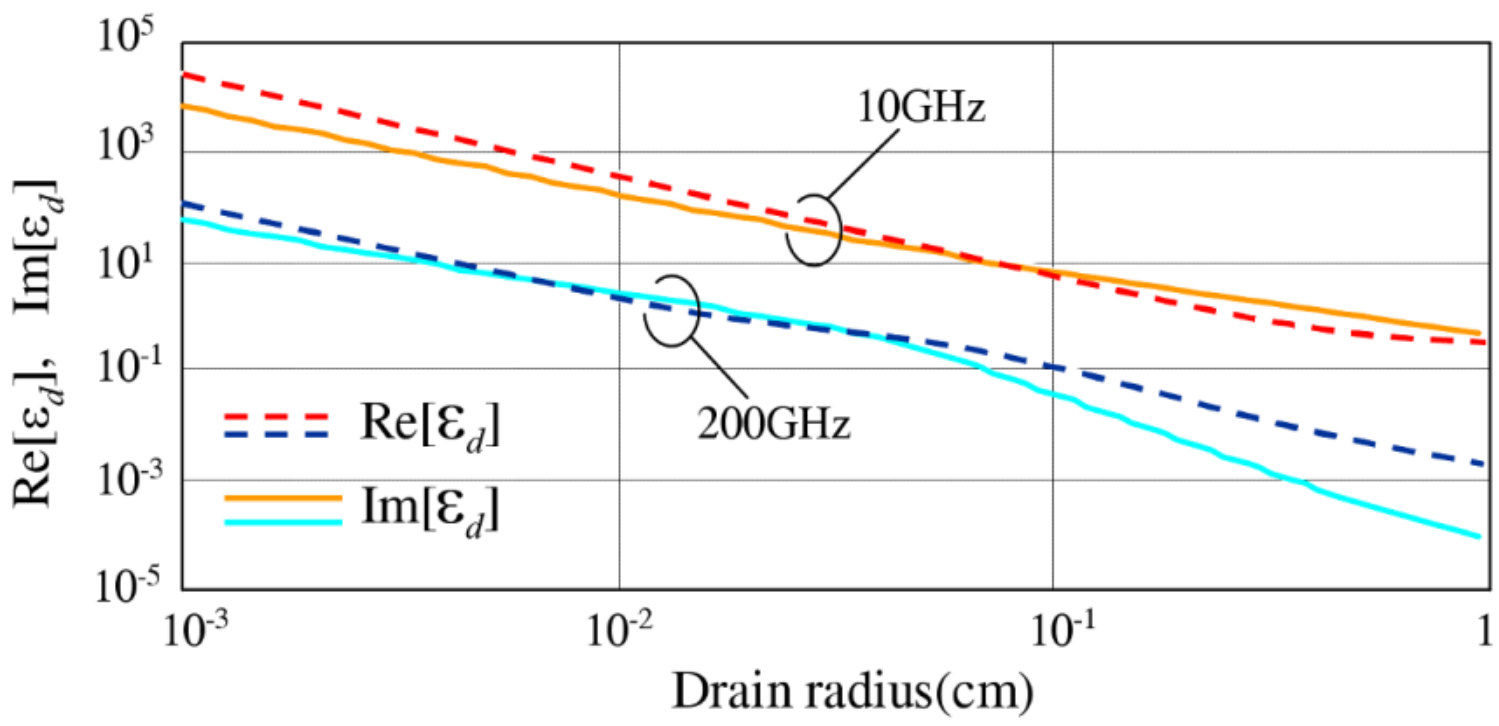

Fig. 3. Dependence of $\varepsilon_d$ (Eq. (14) with drain radius. Dark and bright blue curves correspond to the real and imaginary part for a frequency 200GHz. Red and orange curves to real and imaginary part for a frequency 10GHz.

## 4. Looking inside the drain.

### *4.1. Electric field and current inside the drain.*

The conductivity $\sigma$ of the media inside the drain can be calculated as a function of $\varepsilon_d$ from its definition:

$$\sigma = \frac{\omega}{\mu_0}\mathrm{Im}\left[n_d^2 k_0^2\right] = \frac{4\omega\varepsilon_0}{(1+\rho^2)^2}\mathrm{Im}[\varepsilon_d] \tag{24}$$

Using Eq. (23) we can calculate $\sigma$ as a function of $\nu'$. The electric field inside the drain is given by Eq. (20). Consequently the current density $\mathbf{J}=\sigma\mathbf{E}$ in the drain is:

$$\mathbf{J}(x,y)=\frac{4A\omega\varepsilon_0}{(1+\rho^2)^2}\mathrm{Im}[\varepsilon_d]P_{v'}(\zeta)\mathbf{z} \tag{25}$$

Fig. 4 shows the modulus of the current density inside the drain for a drain radius=0.2 cm and a total current in the source of 1mA when the drain center is at the point (0.519, 0) cm.

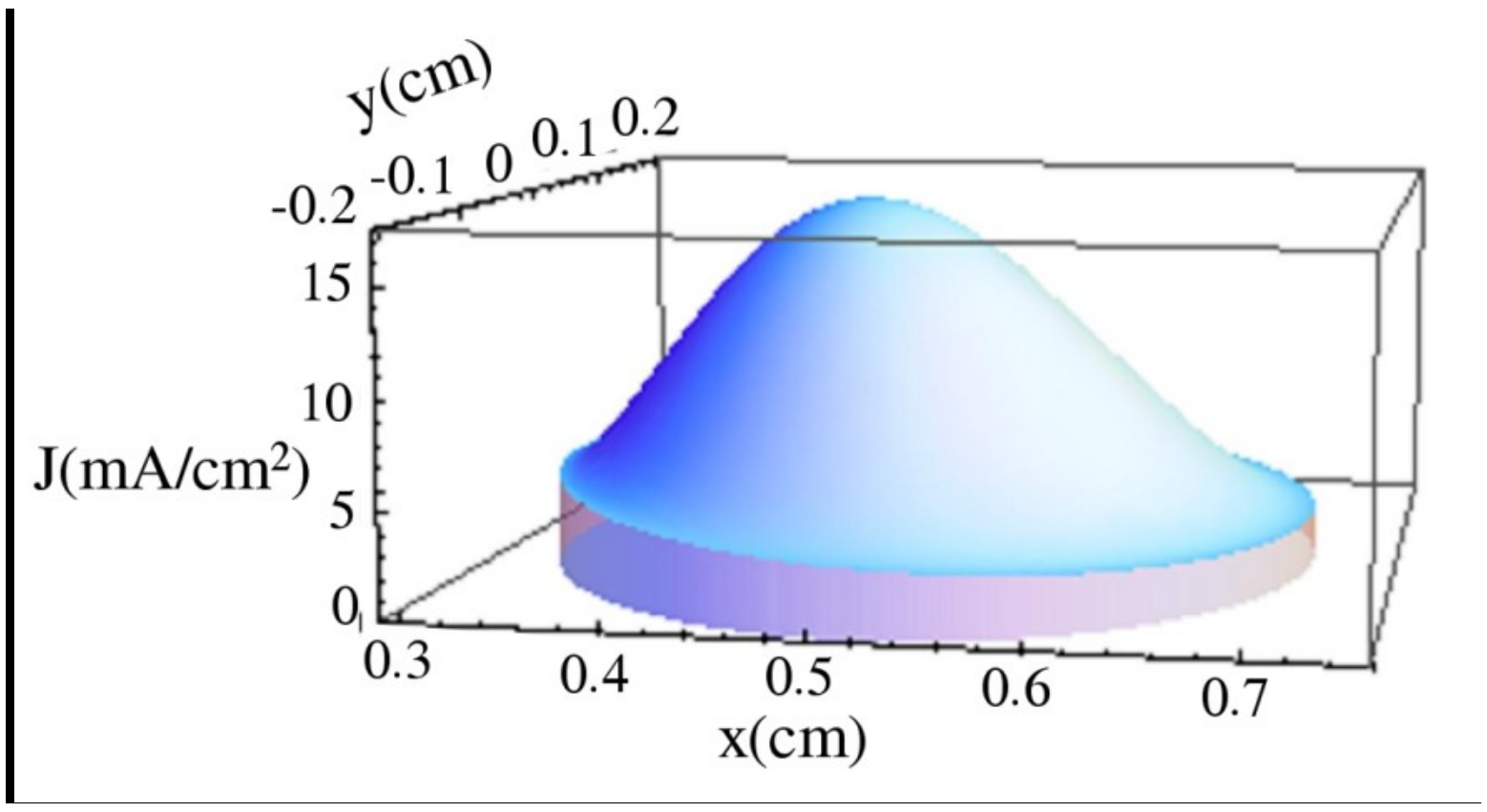

Fig. 4. Representation of the modulus of the current density (mA/cm$^2$) inside the drain. Drain radius=0.2 cm, drain center at (0.519,0) cm. Total current in the source =1mA.

*4.2 Power absorption.*

The power emitted by the source $P$ can be obtained by integrating the Poynting vector over a surface enclosing the source. The MFE is a lossless system, so this power has to equal both the total power entering the drain and the power it absorbs. This integration has been made as shown in Eq. (26). This surface has cylindrical symmetry along the $z$-axis, so for the sake of simplicity we take a surface whose projection on the $x$-$y$ plane is a $\zeta$=constant curve, let's say $\zeta=\zeta_s$. Since the Poynting vector has no $z$-dependence the surface integral is reduced to a line integral along $\zeta=\zeta_s$.

$$P=\frac{1}{2}\mathrm{Re}\left[\int \mathbf{E}\times\mathbf{H}^*\cdot\mathbf{dl}\right]=$$
$$=\frac{1}{2}\mathrm{Re}\left[\left(P_v(\zeta_s)+i\frac{2}{\pi}Q_v(\zeta_s)\right)\left(\frac{1}{i\omega\mu}\left.\frac{d\left(P_v(\zeta)-i\frac{2}{\pi}Q_v(\zeta)\right)}{d\zeta}\right|_{\zeta_s}\right)^*\int_{\zeta=\zeta_s}\left(\frac{\partial\zeta}{\partial y},-\frac{\partial\zeta}{\partial x}\right)\cdot\mathbf{dl}\right] \tag{26}$$

Alternatively, the same power can be calculated as the power absorbed in the volume of the drain, which is:

$$P=\frac{1}{2}\mathrm{Re}\left[\int \mathbf{J}\cdot\mathbf{E}^* dV\right]=\frac{1}{2}\int_{\zeta\le\zeta_d}\sigma\,|AP_{v'}(\zeta)|^2\,dS$$
$$=\frac{1}{2}\mathrm{Re}\left[|A|^2\,P_{v'}(\zeta_d)\left(\frac{1}{i\omega\mu}\left.\frac{dP_{v'}(\zeta)}{d\zeta}\right|_{\zeta_d}\right)^*\right]\int_{\zeta=\zeta_d}\left(\frac{\partial\zeta}{\partial y},-\frac{\partial\zeta}{\partial x}\right)\cdot\mathbf{dl} \tag{27}$$

Fig. 5 shows the power $P$ as function of the frequency for 1mA total current in the point source. Note this implies an amplitude value for Leonhardt's forward wave, correspondingly scaled from Eq. (3) and (8). The power $P$ depends on the source total current and frequency, but obviously it does not depend on the drain radius.

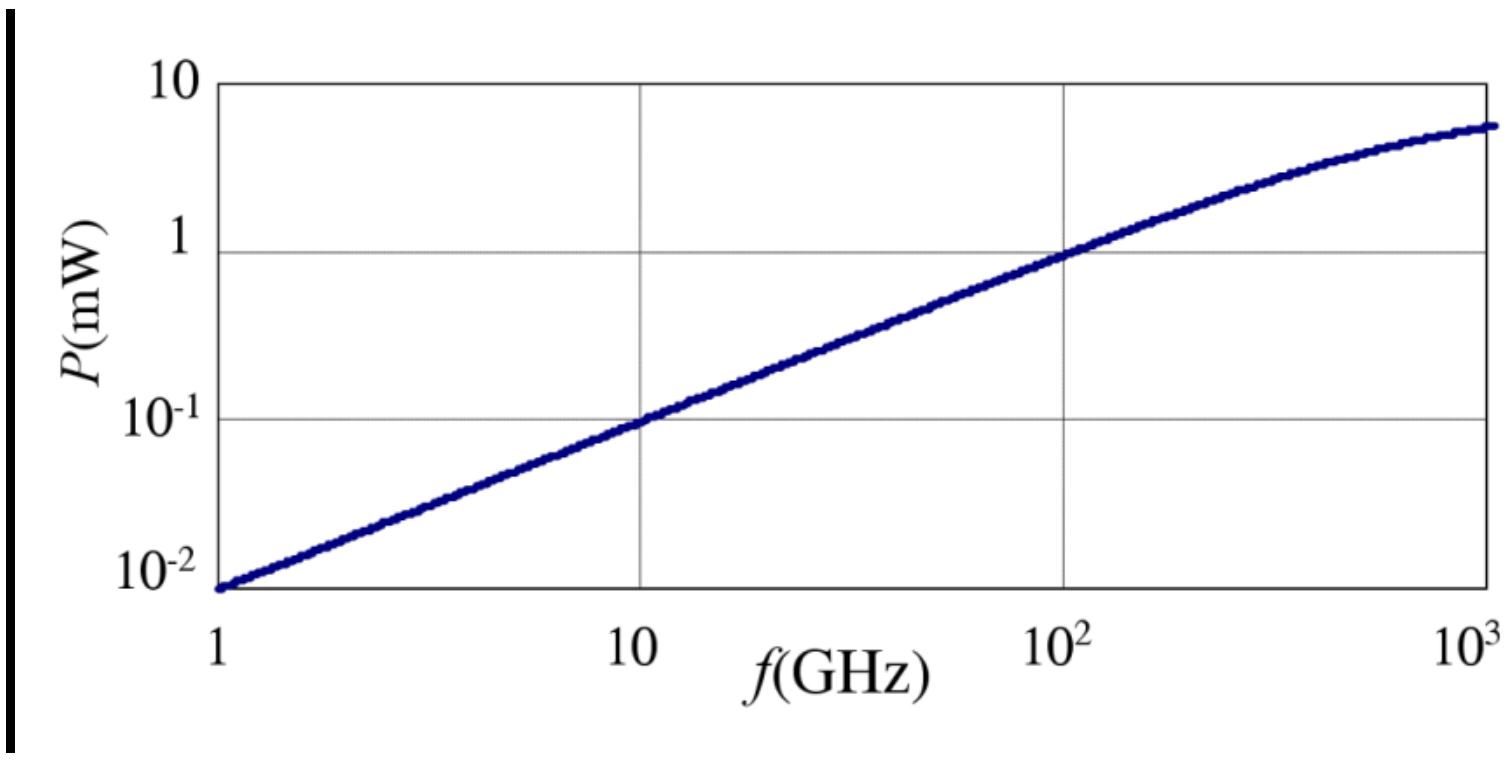

Fig. 5 Power emitted by the source and absorbed by the drain as function of frequency. Total current in the source is 1 mA.

## *5.* Conclusions.

We have found that the perfect drain concept can be modeled as a dissipative region the diameter of which tends to zero. The Leonhardt's forward wave solution inside the MFE can not be obtained for any material used as a drain. We have calculated the complex permittivity of the non-magnetic material forming the perfect drain. When the size of the drain tends to zero, the drain tends to the perfect drain concept Introduced by Leonhardt. Without such a perfect drain, there will be two waves $F_\nu(\zeta)$ and $R_\nu(\zeta)$ inside the MFE (forward and backward respectively, and described in Eq.(31)) as response to the point source.

This concept of the drain as a small dissipative region can be easily included in electromagnetic modeling software. Both characteristics are desirable to analyze and experimentally verify the super-resolution properties of the MFE, *i.e.* to check that the power $P$ changes drastically when the source (or the drain) moves to a neighbor point located at a distance much smaller than the wavelength.

## Appendix. Asymptotic expression when ν >>1, and the Backward wave.

Using Eq. (1) and Eq. (2) in page 162 of [14], functions $P_\nu(\zeta)$ and $Q_\nu(\zeta)$, $\zeta = \cos\theta$, for $\varepsilon < \theta < \pi - \varepsilon$ ($\varepsilon > 0$) can be approximated as:

$$
\begin{aligned}
P_\nu(\cos\theta) &= \frac{\Gamma(\nu)}{\Gamma(\nu+3/2)}\sqrt{\frac{2}{\pi\sin\theta}}\left\{\cos\left((\nu+1/2)\theta\right)+O\left(\nu^{-1}\right)\right\} \\
Q_\nu(\cos\theta) &= \frac{\Gamma(\nu)}{\Gamma(\nu+3/2)}\sqrt{\frac{\pi}{2\sin\theta}}\left\{-\sin\left((\nu+1/2)\theta\right)+O\left(\nu^{-1}\right)\right\}
\end{aligned}
\qquad (28)
$$

Thus, using Eq.(11) we have:

$$
F_\nu(\cos\theta) = \frac{\Gamma(\nu)}{\Gamma(\nu+3/2)}\sqrt{\frac{2}{\pi\sin\theta}}\left\{e^{-i(\nu+1/2)\theta}+o\left(\nu^{-1}\right)\right\} \qquad (29)
$$

which is clearly identified as a wave propagating towards decreasing values of $\theta$ (remember the factor $e^{-i\omega t}$), that is, increasing $\zeta$, from source to drain. This expression is accurate for $\varepsilon<\theta<\pi-\varepsilon$ ($\varepsilon>0$) (that is, not too close to the point source and point drain). Asymptotes to Eq. (28) and (29) suggest to define the function:

$$R_\nu(\zeta)=P_\nu(\zeta)-\frac{2}{\pi}jQ_\nu(\zeta) \tag{30}$$

to describe a field propagating from image point to object point. In analogy to canonical solutions of the Helmholtz Equation (as plan waves, cylindrical waves or spherical waves in free-space), it seems more natural for this MFE problem to define the general solution of Eq. (17) as a superposition of functions $F_\nu$ and $R_\nu$, that is, to rewrite the field E as :

$$E=C\,F_\nu(\zeta)+D\,R_\nu(\zeta) \tag{31}$$

When there is a perfect drain, outside it only the Leonhardt's forward wave $F_\nu$ exists, which implies that $D$ in Eq. (31) is null. Inside the drain, the condition that the field $E$ must be bounded leads to Eq. (20), which in terms of $F_{\nu'}$ and $R_{\nu'}$ implies the condition $C=D$. This means that inside the perfect drain there is a standing wave, which is necessary to avoid the singularity at the image point ($\zeta=1$). This is analogous to the superposition of converging and diverging cylindrical waves, of equal amplitude, in free-space as described by the Hankel functions $H_0^{(1)}(k_0\rho)$ and $H_0^{(2)}(k_0\rho)$, which results in the bounded Bessel function $J_0(k_0\rho)$.

The forward and backward wave defined here for the MFE lens are intimately related to the retarded and advanced field defined in [12] and [13] for the MFE mirror. Using the formulation in [12] and [13], a wave bounded at the image point for the MFE lens can be written:

$$E=\frac{F_{\nu(k_0)}-e^{i\pi(\nu(k_0)-\nu(-k_0))}F_{\nu(-k_0)}}{1-e^{i\pi(\nu(k_0)-\nu(-k_0))}} \tag{32}$$

where the positive and negative wave-numbers are defined as

$$\nu(\pm k_0)=\frac{-1\pm\sqrt{1+4k_0^{\,2}}}{2} \tag{33}$$

Besides the complexity of expression Eq. (32), it can be easily seen that it is (up to a multiplicative constant) simply equal to $P_\nu$ (which is the bounded expression used above in Eq. (20)). This is obtained by direct computation, after considering that $\nu(-k_0)=-\nu(k_0)-1$, and then using Eq. (7) and (16) in page 144 of [14]:

$$\begin{aligned}P_{-\nu-1}(\zeta)&=P_\nu(\zeta)\\ Q_{-\nu-1}(\zeta)&=-\pi\cot(\pi\nu)P_\nu(\zeta)+Q_\nu(\zeta)\end{aligned} \tag{34}$$

**Acknowledgments:**

The authors thank the Spanish Ministry MCEI (Consolider program CSD2008-00066, DEFFIO: TEC2008-03773),) for the support given in the preparation of the present work. The authors also thank Jesús López for creating the video and Bill Parkyn for editing the manuscript.